\documentclass[preprint,aps,draft]{revtex4}

\usepackage{graphicx}% Include figure files
\usepackage{dcolumn}% Align table columns on decimal point
\usepackage{bm}% bold math

\newcommand{\be}{\begin{equation}}
\newcommand{\ee}{\end{equation}}
\newcommand{\ba}{\begin{eqnarray}}
\newcommand{\ea}{\end{eqnarray}}
\newcommand{\bt}{\begin{table}}
\newcommand{\et}{\end{table}}
\newcommand{\brt}{\begin{ruledtabular}}
\newcommand{\ert}{\end{ruledtabular}}
\newcommand{\btu}{\begin{tabular}}
\newcommand{\etu}{\end{tabular}}

\def\la{\langle}
\def\ra{\rangle}

\begin{document}

\preprint{
\noindent
\hfill
\begin{minipage}[t]{3in}
\begin{flushright}
%UPR--T--XXX \\
\vspace*{2cm}
\end{flushright}
\end{minipage}
}

%\draft

\title{Compact Formulas for Tree Amplitudes of Six Partons}
\author{Mingxing Luo and Congkao Wen}
\affiliation{Zhejiang Institute of Modern Physics, Department of Physics,
Zhejiang University, Hangzhou, Zhejiang 310027, P R China}

\date{\today}

\begin{abstract}
Compact results are obtained for tree-level non-MHV amplitudes of six fermions and of four fermions and two gluons,
by using extended BCF/BCFW rules.
Combining with previous results, complete set of tree amplitudes of six partons 
are now available in compact forms. 
\end{abstract}


\maketitle

%\begin{figure}
%\includegraphics[width=3.5in]{y2f.eps}
%\caption{\label{fig:Y2F}Fermion radiative corrections to scalar propagator}
%\end{figure}
%%%%%%%%%%%%%%%%%%%%%%%%%%%%%%%%%%%%%%%%%%%%%%%%%%%%%%%%%%%%%%%%%%%%%%%%%%%%%%%%%%%%%%%%%%%%%
\section{Introduction}

In a period of twelve months or so, tremendous progress has been made in perturbative gauge theories.
It started with a remarkable paper by Ed Witten \cite{witten},
in which a deep relation was pointed out between $N=4$ super gauge theory and one type B topological string theory, 
by re-expressing super gauge theory scattering amplitudes in the language of twistor theories \cite{penrose}.
Particular attention was paid to the so-called maximal helicity violating (MHV) amplitudes,
which can be elegantly expressed in terms of the Parke-Taylor formula \cite{PT,bg1,bg2,mpx}. 
Taking advantage of insights thus gained and by a careful analysis of known helicity amplitudes, 
a novel prescription was proposed to calculate tree level amplitudes (CSW), 
which uses MHV amplitudes as vertices to construct all other amplitudes \cite{csw},

The efficiency of the method is phenomenal and its validity has
been checked by various tree level calculations \cite{wuzhu,khoze,ggk,kosower,wuzhu2,bbk,wuzhu3,gk,zhu}. 
It turns out that the method can be extended to calculate one-loop MHV amplitudes  \cite{bst},
and the paradigm is independent of the large $N_c$ approximation, at least to one-loop \cite{lw1, lw2}. 
The twistor-space structure of one-loop amplitudes are further studied in \cite{csw2,csw3,bern10,cachazo10,bk1,bk2}.
On the other hand, tree-level
amplitudes were also obtained from connected curves in twistor string theories \cite{rsv1,rsv2,rsv3}.

However, the number of MHV diagrams grows rapidly when one includes more external particles.
Real amplitudes may well be much simpler than those suggested by CSW rules,
as evidenced by the extremely simple results obtained for 
NMHV amplitudes by dissolving $N=4$ loop amplitudes into tree ones \cite{bk1,bk2,rsv4}.
Systematically, 
a new set of recursion relations was proposed to calculate tree amplitudes (BCF),
based upon analysis of one-loop amplitudes and infrared relations \cite{rsv4,bcf}.
In \cite{bcfw}, the BCF rules were proved directly by using basic facts of tree diagrams 
with some help from MHV Feynman diagrams (BCFW).
In \cite{bcf}, the recursion relations used two adjacent gluons of opposite helicity as reference gluons.
Actually,  reference gluons do not have to be adjacent and they can also be of the same helicity \cite{bcfw}.
In \cite{lw3}, these recursion relations are extended to include fermions.
From which, MHV and $\overline{\rm MHV}$ amplitudes, 
and non-MHV amplitudes of processes with two fermions and four gluons are reproduced correctly. 
On the other hand, these recursion relations may also help to 
determine rational functions appearing in one-loop QCD amplitudes \cite{bdk05}.

In this paper, we will calculate amplitudes of six partons, 
by using these new recursion relations and by extended CSW rules \cite{khoze}.
These include non-MHV amplitudes of six fermions and of four fermions and two gluons,
with fermions in either adjoint representations (gauginos)
or fundamental representations (quarks) of the gauge group.
Results obtained from these two methods are equivalent to each other,
and equivalent to those obtained by conventional field theory methods \cite{gk1,xzc,gk2,sw}.
In particular, our results assume extremely simple and compact forms,
considering the number of ordinary Feynman diagrams involved.
Combining with previous results in \cite{lw3}, 
complete set of all tree amplitudes of six partons 
are now available in compact forms. 

Following this introduction, we will present a brief review of the BCE/BCFW rules and extensions in section 2.
In section 3, we will list amplitudes of six partons.
And we will conclude in section 4.

\section{Review of the BCF/BCFW Approach and Extensions}
Now we give a brief review of the BCF/BCFW approach.
Take an $n$-gluon tree-level amplitude of any helicity configuration. 
As amplitudes of gluons with all the same helicity vanish,
we can always arrange gluons such that the $(n-1)$-th gluon has negative helicity
and the $n$-th gluon has positive helicity. These two lines will be taken as reference lines.
Labeling external particles by $i$, the following recursion relation was claimed in \cite{bcf}:
\ba
A_n(1,2,\ldots , (n-1)^-,n^+) & =& \sum_{i=1}^{n-3}\sum_{h=+,-} 
\left( A_{i+2}(\hat{n},1,2,\ldots, i,-\hat{P}^h_{n,i} ) \right. \label{eq2.1}  \\
&& \times {1\over P^2_{n,i}} \left. A_{n-i}(+\hat{P}^{-h}_{n,i}, i+1,\ldots , n-2, {\hat{n-1} } ) \right) \nonumber
\ea
where
\ba
P_{n,i} & = & p_n+p_1+\ldots + p_i, \nonumber \\ 
\hat{P}_{n,i} & =& P_{n,i} +{P_{n,i}^2\over \langle n-1|P_{n,i}|n]} \lambda_{n-1} \tilde\lambda_{n}, \nonumber\\ 
\hat p_{n-1} & = & p_{n-1} -{P_{n,i}^2\over \langle n-1|P_{n,i}|n]} \lambda_{n-1}\tilde\lambda_{n} , \\
\hat p_{n} & = & p_{n} +{P_{n,i}^2\over\langle n-1|P_{n,i}|n]} \lambda_{n-1} \tilde\lambda_{n}. \nonumber
\ea
The formula has a natural meaning in $(--++)$ signature.
Notice that ${\hat P}_{n,i}^2 = {\hat p}^2_n = {\hat p}^2_{n-1} = 0$, 
so each tree-level amplitude in Eq (\ref{eq2.1}) has all external gluons on-shell. 
Still, energy-momentum conservation is preserved.
In \cite{bcfw}, 
it has been shown that reference gluons do not have to be adjacent and they can be of the same helicity.

To streamline notations, it is expedient to define 
\be
\begin{array}{rcl}
K_i^{[r]} &\equiv& p_i+p_{i+1}+\cdots+p_{i+r-1} \\ %, \nonumber \\
t_i^{[r]} &\equiv& \left( K_i^{[r]} \right)^2  % \nonumber \\
\end{array}
\ee
And the following relations will be useful to make simplifications:
\be
\langle A | \hat{P}_{n,i} \rangle  = -{1\over \omega} \langle A | P_{n,i} |n], \ \ \ \
\left[ \right. \hat{P}_{n,i} | B ]  = -{1\over\bar{\omega}} \langle n-1 | P_{n,i} | B ]
\ee
Factors $\omega$ and $\bar{\omega}$ always show up in final results
in the combination of $\omega\bar{\omega}$, which is equal to $\la n-1 | P_{n,i} | n\ra$.
This completes the BCF/BCFW prescription.

These rules can be extended to include fermions.
Specifically, one chooses any two external lines, either gluons or fermions, as reference lines;
then shifts relevant momenta exactly in the same manner as those in Eq. (2) and finally,
combines sub-amplitudes of less numbers of external lines together in the same manner as that in Eq. (1).
Of course, there could now be diagrams in which sub-diagrams are linked by internal fermionic lines,
but their propagators are the same as those of gluons.
But there are some qualifications, 
which are closely related to behaviors of meromorphic functions $A(z)$ defined in Eq (2.3) of \cite{bcfw}.
Specifically, one cannot take two adjacent fermion lines of opposite helicities as reference lines,
since in such cases, 
\be
A_{\rm MHV} (\Lambda_1^- \Lambda_2^+ g_i^-)\sim
\frac{\langle i \hat{1}\rangle ^3\langle i \hat{2} \rangle }
{\langle n \hat{1} \rangle \langle \hat{1} \hat{2} \rangle \langle \hat{2} 3 \rangle}
\sim 1.
\ee
On the other hand, one fermion and an adjacent gluon of the same helicity cannot be taken as reference lines either, as 
\be
A_{\rm MHV}(\Lambda_1^- g_2^- \Lambda_i^+) \sim 1. 
\ee
By carefully avoiding these particular cases, 
these extensions were justified in \cite{lw3} by reasonings paralleling those in \cite{bcfw}.

\section{Scattering Amplitudes of Six Partons}
In \cite{lw3}, MHV and $\overline{\rm MHV}$ amplitudes with fermions, 
and non-MHV amplitudes of processes with two fermions and four gluons are reproduced correctly
by using the extended BCF/BCFW rules. 
We now apply these rules to calculate other amplitudes of six partons.
These include non-MHV amplitudes of six fermions and of four fermions and two gluons.
We start with fermions in adjoint representations (gauginos).
The results for fermions in fundamental representations (quarks)
can be obtained from those of gauginos by making proper identifications.
We have also calculated these amplitudes by extended CSW rules \cite{csw,khoze},
and obtained results totally equivalent to the ones given above.
On the other hand, these results are equivalent to the ones obtained by conventional field theory calculations
\cite{gk1,xzc,gk2,sw}.
For cases of four fermions and two gluons, we have chosen two ``good" reference lines,
so the validity of recursion relations can be proved by using basic facts of tree diagrams and
without the help of CSW rules \cite{lw3}.
Direct comparisons with results in \cite{gk1,xzc} have thus not been performed.

\subsection{One flavor fermions in the adjoint representation}
We start with one flavor gauginos.
The color factor ${\rm Tr} (T^{a_1} \cdots T^{a_6})$ will be stripped off 
and only the kinematic factors will be listed.
First, six fermions.
\ba
A(\Lambda_1^- \Lambda_2^- \Lambda_3^- \Lambda_4^+ \Lambda_5^+ \Lambda_6^+)
&=&\frac{\langle 1|2+3|4]^2}{t_2^{[3]}[3 4]\langle 6 1 \rangle \langle 5|4+3|2]} \nonumber \\
&-&\frac{[6|1+2|3\rangle ^2}{t_3^{[3]}[1 6]\langle 3 4 \rangle \langle 5|6+1|2]}
\ea
\ba
A(\Lambda_1^+ \Lambda_2^+ \Lambda_3^- \Lambda_4^- \Lambda_5^+ \Lambda_6^-)
&=& \frac{[4 6]\langle 3|2+1|5]^3}{t_1^{[3]}\langle 2 3 \rangle [4 5][5 6]\langle 3|1+2|6]\langle 1|2+3|4]}
\nonumber \\
&+&\frac{\langle 1 5 \rangle \langle 6|3+4|2]^3}
{t_2^{[3]}[2 3]\langle 5 6 \rangle \langle 6 1 \rangle \langle 5|4+3|2]\langle 1|2+3|4]} \\
&+&\frac{[1 2]^2\langle 3 4 \rangle^3 \langle 5|1+2|6]}
{t_3^{[3]}[1 6]\langle 3 4 \rangle \langle 4 5 \rangle \langle 5|6+1|2]\langle 3|2+1|6] } \nonumber
\ea
All other amplitudes of six gauginos of one-flavor can be obtained from these two formulas,
by complex conjugations and by using generalized dual Ward identities \cite{mpx},
\be
A(1, 2, 3, \cdots, n) \pm A(2,1,3, \cdots, n) \pm A(2,3,1,\cdots, n) \pm \cdots \pm A(2,3,\cdots, 1, n)=0
\label{ward}
\ee
where the minus sign is for cases when $2, \cdots, 1, \cdots, n$ are 
odd permutations of $1, 2, \cdots, n$ among fermions and the plus sign is for all other cases. For example, 
\ba
A(\Lambda_1^- \Lambda_2^+ \Lambda_3^- \Lambda_4^+ \Lambda_5^- \Lambda_6^+)
&=& +A(\Lambda_2^+ \Lambda_1^- \Lambda_3^- \Lambda_4^+ \Lambda_5^- \Lambda_6^+)
-A(\Lambda_2^+ \Lambda_3^- \Lambda_1^- \Lambda_4^+ \Lambda_5^- \Lambda_6^+) \nonumber \\
&&+A(\Lambda_2^+ \Lambda_3^- \Lambda_4^+ \Lambda_1^- \Lambda_5^- \Lambda_6^+)
-A(\Lambda_2^+ \Lambda_3^- \Lambda_4^+ \Lambda_5^- \Lambda_1^- \Lambda_6^+)
\ea
This relation and similar ones have been checked explicitly.

Now we list amplitudes of four gauginos and two gluons.
\ba %1
A(g_1^- g_2^+ \Lambda_3^- \Lambda_4^+ \Lambda_5^- \Lambda_6^+)
&=& \frac{\langle 1 3 \rangle ^3[4 6]^3[5|3+2|1\rangle }
{t_1^{[3]}\langle 1 2 \rangle \langle 2 3 \rangle [4 5][5 6][6|1+2|3\rangle [4|3+2|1\rangle } \nonumber \\
&+&\frac{[2 4]^3\langle 1 5 \rangle ^3}{t_2^{[3]}\langle 5 6 \rangle [3 4][2|3+4|5\rangle [4|3+2|1\rangle } 
\label{eq:11}\\
&+&\frac{[2 6]^3\langle 3 5 \rangle ^3[2|1+6|4\rangle}
{t_3^{[3]}[1 2][1 6]\langle 3 4 \rangle \langle 4 5 \rangle[6|1+2|3\rangle [2|1+6|5\rangle }\nonumber
\ea
\ba %2
A(g_1^- g_2^+ \Lambda_3^- \Lambda_4^- \Lambda_5^+ \Lambda_6^+)
&=&-\frac{\langle 1 3 \rangle ^3[5 6]^2}{t_1^{[3]}\langle 1 2 \rangle \langle 2 3 \rangle [4 5][6|1+2|3\rangle }
\nonumber \\
&+&\frac{[6 2]^3\langle 3 4 \rangle ^2 }{t_3^{[3]}[1 2][1 6]\langle 4 5 \rangle [6|1+2|3\rangle }
\label{eq:12}
\ea
\ba %3
A(g_1^- g_2^+ \Lambda_3^- \Lambda_4^+ \Lambda_5^+ \Lambda_6^-)
&=& -\frac{\langle 1 3 \rangle ^3[4 5]^2\langle 1|2+3|6]}
{t_1^{[3]}\langle 1 2 \rangle \langle 2 3 \rangle [5 6]\langle 1|2+3|4][6|1+2|3\rangle } \nonumber \\
&-&\frac{\langle 1 6 \rangle ^2\langle 1 5 \rangle [2 4]^3}
{t_2^{[3]}[3 4]\langle 5 6 \rangle \langle 5|4+3|2]\langle 1|2+3|4]} 
\label{eq:13}
\\
&+&\frac{[6 2]\langle 3|1+6|2]^3}
{t_3^{[3]}[1 2][1 6]\langle 3 4 \rangle \langle 5|1+6|2][6|1+2|3\rangle} \nonumber
\ea
\ba %4
A(g_1^- g_2^+ \Lambda_3^+ \Lambda_4^- \Lambda_5^- \Lambda_6^+)
&=& -\frac{\langle 1 3 \rangle \langle 1|2+3|6]^3}
{t_1^{[3]}\langle 1 2 \rangle \langle 2 3 \rangle[5 6][6|1+2|3\rangle \langle 1|2+3|4]} \nonumber \\
&-&\frac{\langle 1 5 \rangle ^3[2 3]^2[2 4]}
{t_2^{[3]}\langle 5 6 \rangle [3 4]\langle 5|4+3|2]\langle 1|2+3|4]} 
\label{eq:14} \\
&+&\frac{[6 2]^3\langle 4 5 \rangle ^2\langle 3|6+1|2]}
 {t_3^{[3]}[1 2][1 6] \langle 3 4\rangle \langle 5|6+1|2][6|1+2|3\rangle } \nonumber
\ea
\ba %5
A(g_1^- g_2^+ \Lambda_3^+ \Lambda_4^- \Lambda_5^+ \Lambda_6^-)
&=& \frac{\langle 1 3 \rangle [4 6]\langle 1|2+3|5]^3}
{t_1^{[3]}\langle 1 2 \rangle \langle 2 3\rangle [4 5][5 6][6|1+2|3\rangle \langle 1|2+3|4]} \nonumber \\
&+&\frac{\langle 1 6 \rangle ^2 \langle 1 5\rangle [2 3]^2[2 4]}
{t_2^{[3]}[3 4]\langle 5 6\rangle \langle 5|4+3|2]\langle 1|2+3|4]} 
\label{eq:15}\\
&-&\frac{\langle 5 3 \rangle [2 6]\langle 4|6+1|2]^3}
{t_3^{[3]}[1 2][1 6]\langle 3 4\rangle \langle 4 5\rangle \langle 5|6+1|2][6|1+2|3\rangle } \nonumber
\ea
\ba %6
A(g_1^- g_2^+ \Lambda_3^+ \Lambda_4^+ \Lambda_5^- \Lambda_6^-)
&=& -\frac{\langle 1 3 \rangle \langle 1|2+3|4]^2}
{t_1^{[3]}\langle 1 2\rangle \langle 2 3\rangle [4 5][6|1+2|3\rangle } \nonumber \\
&-&\frac{[2 6]\langle 5|6+1|2]^2}{t_3^{[3]}[1 2][1 6]\langle 4 5\rangle [6|1+2|3\rangle }
\label{eq:16}
\ea
Again, all other amplitudes of four fermions can be obtained from these six formulas, 
by complex conjugations and by using generalized dual Ward identities Eq. (\ref{ward}).

\subsection{Multi-flavor fermions in adjoint representations}
Now, multi-flavor fermions in adjoint representations. For six fermions of three flavors,
\ba
A(\alpha_1^- \beta_2^+ \beta_3^- \gamma_4^+ \gamma_5^- \alpha_6^+)
&=&\frac{\langle 1 3\rangle^2[4 6]^2 \langle 3|1+2|4]}{t_1^{[3]}\langle 2 3\rangle [4 5]\langle 1|2+3|4]\langle 3|2+1|6]}
\nonumber \\
&&-\frac{\langle 1 5\rangle^2[2 4]^2 \langle 1|3+4|2]}{t_2^{[3]}\langle 6 1\rangle [2 3]\langle 1|2+3|4]\langle 5|4+3|2]}
\\
&&-\frac{\langle 3 5\rangle^2[2 6]^2 \langle 5|1+2|6]}{t_3^{[3]}\langle 4 5\rangle [6 1]\langle 3|1+2|6]\langle 5|6+1|2]}
\nonumber
\ea
\ba
A(\alpha_1^- \beta_2^- \beta_3^+ \gamma_4^+ \gamma_5^- \alpha_6^+)
&=&-\frac{\langle 1 2\rangle^2[4 6]^2 \langle 3|1+2|4]}{t_1^{[3]}\langle 2 3\rangle [4 5]\langle 1|2+3|4]\langle 3|2+1|6]}
\nonumber \\
&&+\frac{\langle 1 5\rangle^2[3 4]^2 \langle 1|3+4|2]}{t_2^{[3]}\langle 6 1\rangle [2 3]\langle 1|2+3|4]\langle 5|4+3|2]}
\\
&&+\frac{\langle 5|1+2|6]^3}{t_3^{[3]}\langle 4 5\rangle [6 1]\langle 3|1+2|6]\langle 5|6+1|2]}
\nonumber 
\ea
For six fermions of two flavors,
\ba
A(\eta_1^- \eta_2^+ \Lambda_3^- \Lambda_4^- \Lambda_5^+ \Lambda_6^+)
&=&\frac{\langle 1 3\rangle^2[5 6]^2}{t_1^{[3]}\langle 1 2\rangle [4 5]\langle 3|4+5|6]}
\nonumber \\
&&-\frac{\langle 3 4\rangle^2[2 6]^2}{t_3^{[3]}[1 2]\langle 4 5\rangle \langle 3|4+5|6]}
\ea
\ba
A(\eta_1^- \eta_2^+ \Lambda_3^+ \Lambda_4^+ \Lambda_5^- \Lambda_6^-)
&=&\frac{\langle 1|2+3|4]^2}{t_1^{[3]}\langle 1 2\rangle [4 5]\langle 3|4+5|6]}
\nonumber \\
&&-\frac{[2|3+4|5\rangle^2}{t_3^{[3]}[1 2]\langle 4 5\rangle \langle 3|4+5|6]}
\ea
\ba
A(\eta_1^- \eta_2^+ \Lambda_3^- \Lambda_4^+ \Lambda_5^- \Lambda_6^+)
&=&-\frac{\langle 1 3\rangle^2[4 6]^3 \langle 1|2+3|5]}
{t_1^{[3]}\langle 1 2\rangle [4 5][5 6]\langle 1|2+3|4]\langle 3|4+5|6]}
\nonumber \\
&&-\frac{\langle 1 5\rangle^2[2 4]^2 \langle 5|6+1|4]}
{t_2^{[3]}[3 4]\langle 5 6\rangle \langle 1|2+3|4]\langle 5|6+1|2]}
\\
&&+\frac{[2 6]^2\langle 3 5\rangle^3 \langle 4|6+1|2]}
{t_3^{[3]}[1 2]\langle 3 4\rangle \langle 4 5\rangle \langle 3|4+5|6]\langle 5|6+1|2]}
\nonumber 
\ea
For four fermions of two flavors and two gluons
\ba
A(\eta^+_1 g^-_2 g^+_3 \Lambda^-_4 \Lambda^+_5 \eta^-_6)
&=& -\frac{\langle 4 6\rangle^2[1 3]^3 \langle 4|1+2|3]} 
{t_1^{[3]}[1 2][2 3]\langle 4 5\rangle \langle 6|1+2|3][1|2+3 | 4 \rangle}
\nonumber \\
&&-\frac{\langle 2 4\rangle^3[1 5]^2 \langle 2|3+4|1]}
{t_2^{[3]}\langle 2 3\rangle \langle 3 4\rangle [6 1][1|2+3 |4\rangle \langle 2|3+4|5]}
\\
&&-\frac{\langle 2 6\rangle^3[3 5]^3}
{t_3^{[3]}[4 5]\langle 6 1\rangle \langle 2|3+4|5]\langle 6|5+4|3]}
\nonumber 
\ea
\ba
A(\eta^-_1 g^-_2 g^+_3 \Lambda^+_4 \Lambda^-_5 \eta^+_6)
&=&-\frac{[1 3]\langle 4|1+2|3]\langle 5|1+2|3]^2}
{t_1^{[3]}[1 2][2 3]\langle 4 5\rangle [1|2+3|4\rangle \langle 6|1+2|3]}
\nonumber \\
&&-\frac{\langle 2 4\rangle \langle 2|3+4|1]\langle 2|3+4|6]^2}
{t_2^{[3]}\langle 2 3\rangle \langle 3 4\rangle [6 1]\langle 2|3+4|5][1|2+3|4\rangle}
\\
&&-\frac{[3 5][3 4]^2\langle 2 6\rangle \langle 1 2\rangle^2}
{t_3^{[3]}[4 5]\langle 6 1\rangle \langle 2|3+4|5]\langle 6|5+4|3]}
\nonumber 
\ea
\ba
A(\Lambda^+_1 g^-_2 g_3^+ \Lambda^-_4 \eta^+_5 \eta^-_6)
&=&-\frac{\langle 4 6\rangle^2 [1 3]^3}
{t_1^{[3]}[1 2][2 3]\langle 5 6\rangle [1|2+3|4 \rangle}
\nonumber \\
&&-\frac{\langle 2 4\rangle^3[1 5]^2}
{t_2^{[3]}\langle 2 3\rangle \langle 3 4\rangle[5 6][1|2+3|4\rangle}
\ea
\ba
A(\Lambda^-_1 g^-_2 g_3^+ \Lambda^+_4 \eta^-_5 \eta^+_6)
&=&-\frac{[1 3]\langle 5|1+2|3]^2}
{t_1^{[3]}[1 2][2 3]\langle 5 6\rangle [1|2+3|4\rangle}
\nonumber \\
&&-\frac{\langle 2 4\rangle \langle 2|3+4|6]^2}
{t_2^{[3]}\langle 2 3\rangle \langle 3 4\rangle[5 6][1|2+3|4\rangle}
\ea
Up to an overall minus sign,
$A(\eta^-_1 g^-_2 g^+_3 \Lambda^-_4 \Lambda^+_5 \eta^+_6)$, 
$A(\eta^+_1 g^-_2 g^+_3 \Lambda^+_4 \Lambda^-_5 \eta^-_6)$,
$A(\Lambda^+_1 g^-_2 g_3^+ \Lambda^-_4 \eta^-_5 \eta^+_6)$ and
$A(\Lambda^-_1 g^-_2 g_3^+ \Lambda^+_4 \eta^+_5 \eta^-_6)$ 
can be obtained from Eqs. (\ref{eq:13}), (\ref{eq:14}), (\ref{eq:12}) and (\ref{eq:16}), respectively,
by shifting the indices $i \rightarrow i+1$.
All other cases can be obtained from these 
by dual Ward identities Eq. (\ref{ward}) and/or complex conjugations.
%added

\subsection{Fermions in fundamental representations}
%added
In quantum chromodynamics, quarks are in fundamental representations of $SU(N)$ gauge groups.
Still, tree diagrams involving quarks are factorized property.
Specifically, for a fixed color ordering $\sigma$, the amplitude
with $m$ quark-anti-quark pairs and $l$ gluons is still a perfect product,
\be
T_{l+2m} (\{c_{\sigma(i)}\}) \ A_{l+2m} (\{k_{\sigma(i)},h_{\sigma(i)}\}) \, ,
\label{three}
\ee
where
\be
T_{l+2m} \, = \,
{(-1)^p\over N^p} (T^{a_1} \ldots T^{a_{l_1}})_{i_1 \alpha_1}
(T^{a_{l_1+1}} \ldots T^{a_{l_2}})_{i_2 \alpha_2} \ldots
(T^{a_{l_{m-1}+1}} \ldots T^{a_{l}})_{i_m \alpha_m} \, .
\label{cfqq}
\ee
Here $\{ l_k \}$ correspond to an arbitrary partition of an arbitrary permutation of the $l$ gluon indices; 
$\{i_k \}$ are color indices of quarks, and $\{ \alpha_k\}$ of anti-quarks.
Each external quark is connected by a fermion line to an external anti-quark (all particles are counted as incoming). 
When quark $i_k$ is connected by a fermion line to anti-quark $\alpha_k$, we set $\alpha_k=\bar{i_k}$. 
The set of $\{ \alpha_k\}$  is then a permutation of the set $\{ \bar{i_k} \}$.
Finally, the power $p$ is equal to the number of times $\alpha_k=\bar{i_k}$.
For $\{\alpha_k\} = \{\bar{i_k}\}$, $p=m-1$.

With the color information being stripped off, 
the kinematic amplitudes $A_{l+2m}$ in Eq. (\ref{three}) do not distinguish between quarks and gluinos.
Up to possible overall minus signs,
they can be obtained from the results of fermions in adjoint representations by proper identifications.
For example, in cases of four fermions, the leading term proportional to
$ (\{T^A\})_{i_1, \bar{i}_2}  (\{T^B\})_{i_2, \bar{i}_1}$
can be obtained from
\be
A(\Lambda_{i_1} \{g_A\} \eta_{\bar{i}_2} \eta_{i_2} \{g_B\} \Lambda_{\bar{i}_1})
\ee
where $\{T^A\} = T^{a_1} T^{a_2} \cdots$ and $\{g_A\} = g_{a_1} g_{a_2} \cdots$,
and similarly for $\{T^B\}$ and $\{g_B\}$.
The sub-leading term proportional to $(1/N) (\{T^A\})_{i_1, \bar{i}_1}  (\{T^B\})_{i_2, \bar{i}_2}$
can obtained from
\be
A(\Lambda_{i_1} \{g_A\} \Lambda_{\bar{i}_1} \eta_{i_2} \{g_B\} \eta_{\bar{i}_2})
\ee
In cases of six fermions, the leading term proportional to
$ (\{T^A\})_{i_1, \bar{i}_3} (\{T^B\})_{i_3, \bar{i}_2} (\{T^C\})_{i_2, \bar{i}_1}$
can obtained from
\be
A(\alpha_{i_1} \{g_A\} \gamma_{\bar{i}_3} \gamma_{i_3} \{g_B\} \beta_{\bar{i}_2} \beta_{i_2} \{g_C\} \alpha_{\bar{i}_1})
\ee
The sub-leading term proportional to
$(1/N) (\{T^A\})_{i_1, \bar{i}_1} (\{T^B\})_{i_2, \bar{i}_3} (\{T^C\})_{i_3, \bar{i}_2}$
can obtained from
\be
A(\alpha_{i_1} \{g_A\} \alpha_{\bar{i}_1} \beta_{i_2} \{g_B\} \gamma_{\bar{i}_3} \gamma_{i_3} \{g_C\} \beta_{\bar{i}_2})
+ A(\alpha_{i_1} \{g_A\} \alpha_{\bar{i}_1} \gamma_{i_3} \{g_C\} \beta_{\bar{i}_2} \beta_{i_2} \{g_B\} \gamma_{\bar{i}_3} )
\label{sub2}
\ee
Notice the presence of two terms in Eq. (\ref{sub2}), as they are both present and inequivalent,
which is actually a generic feature for sub-leading contributions.
Generalizations to general cases are straightforward.
When quarks are of the same flavor, there will be $u$-channel contributions which can be constructed
in the same way.
As concrete applications of these rules, 
all amplitudes of six partons involving quarks can be obtained from the results in the previous subsection.

\section{Conclusion}
In this paper, we have calculated amplitudes of six partons in spinor gauge theories,
with fermions in either adjoint representations
or fundamental representations of gauge groups,
by using extended BCF/BFCW rules and by using extended CSW prescriptions.
Results obtained by these two methods are equivalent to each other,
and equivalent to those by conventional field theory calculations.
As expected, extremely compact formulas were obtained and computations are much simpler.
Combining with previous results in \cite{lw3}, 
complete set of all tree amplitudes of six partons are now available in compact forms. 
Naturally, similar calculations are expected to be performed for more complex situations,
so these recursion rules will serve as powerful tools in the analysis of jet physics.

\begin{acknowledgments} 
We thank B. Cai and F. Xu for help on computer matters.
This work is supported in part by the National Science Foundation of China (10425525).
\end{acknowledgments}

\end{document}